\documentclass[titlepage,12pt]{article}
\usepackage{amssymb}
\usepackage{amsfonts}
\usepackage{amsmath}
\begin{document}

{\bf \Large Mathematical Models in Physics : a Quest \\ \\ for Clarity} \\ \\

{\bf Elemer E Rosinger} \\
Department of Mathematics \\
and Applied Mathematics \\
University of Pretoria \\
Pretoria \\
0002 South Africa \\
eerosinger@hotmail.com \\ \\

{\bf Abstract} \\

The role of mathematical models in physics has for longer been well established. The issue of
their proper building and use appears to be less clear. Examples in this regard from
relativity and quantum mechanics are mentioned. Comments concerning a more appropriate way in
setting up and using mathematical models in physics are presented. \\ \\

{\bf 1. A Quest for Clarity ...} \\

As it happens, mathematics evokes rather strong feelings among the variety of
non-mathematicians who use it nowadays. Among them, here we shall focus on theoretical
physicists. \\

On one hand, we have in [W] a rather extraordinary point of view expressed by one of the old
guard of Nobel laureate quantum physicists. On the other, and rather more typically, many a
physicist may see mathematics in a way similar to having to go to the dentist ... \\

The mentioned view of Wigner had at the time, and ever since, received few reactions or
comments. And none of them seemed to reach deep enough towards a better understanding of the
alleged unreasonable effectiveness. As for the other and far more frequent view of mathematics
among physicists, it is but a rather natural and thus an easy to understand subjective
reaction of anyone whose interest is definitely not in mathematics as such, but only in its
use, if an when unavoidable. \\

Let us, therefore, start by trying to clarify what is so different between mathematics and
physics, and what makes the former so unavoidable in the pursuit of latter, and does so at
least on occasions. \\

Clearly, both mathematics and physics, as much as chemistry and a number of branches of
engineering or biology, among others, are supposed to be {\it precise}. And they certainly are
so in a sense in which, for instance, philosophy, psychology or sociology are not, and are not
even expected to be. \\

Above that, however, mathematics has a second essential feature, not exhibited by any other
sciences, namely, it is {\it abstract}. And this abstract feature gives mathematics one of its
unique powers, namely, its {\it generality}. Indeed, the rather elementary mathematical
formula 1 + 1 = 2, for instance, is not only about, say, one atom of hydrogen added to another
similar atom, and then giving two such atoms, just as much as it is not only about one apple
plus another apple, making altogether two apples. Instead, due precisely to the {\it abstract}
nature even of the rather elementary concept of positive integer, such as 1 and 2, among
others, the above formula has such an immense {\it generality}, as to apply to no less than an
{\it infinity} of specific instances. And in fact, that infinity is so immense, as to be
inconceivable even mathematically ... \\
In this regard, physics, for instance, does not much need to concern itself with an
inconceivable infinity of, say, hydrogen atoms, just like biology seems in no urgent need to
deal with an infinity of, say, living cells ... \\

In this way, it is rather its abstract nature, thus its generality, than merely its precision,
which makes mathematics useful, if not in fact necessary and inevitable, in a variety of other
scientific pursuits. Indeed, none of the other sciences have such an arsenal of concepts,
methods and results which, due to their immense generality resulting from their abstract
nature, may be applicable across all other sciences. \\
The only comment in this regard one may make is that, as far as other sciences are concerned,
the generality of mathematics, giving its ability to deal with inconceivable infinities of
specific instances, is more than seems to be practically needed. \\
On the other hand, and as well known in mathematics, dealing with infinities, be they
conceivable or not, is at least since the introduction of Cantor's set theory in the late
1800s, a rather natural internal mathematical affair ... \\

And thus, one of the main issues related to the role of mathematical modeling in physics, and
more generally, in sciences, is that, so far, we humans have not developed any other
comparatively precise and abstract, hence generally valid, science as is the case with
mathematics. \\

A second issue in this context is quite natural as well, even if it often leads to lots of
confusion. And such confusions are, needless to say, not limited to physics alone, even if
here we shall only focus on the way mathematical models are set up and used in physics. \\
Namely, a mathematical model is of course supposed to incorporate two rather different
ingredients : the insights of physics, and the mathematical structures available, or rather,
known to those who set up the respective specific mathematical model. \\

Here, naturally, two phenomena can occur : the mathematical model will in due time undergo
changes, mostly due to theoretical and experimental considerations in physics, and on less
frequent occasions, new mathematics may be developed, as needed in, or even merely inspired
by, the specific mathematical modelling process at hand. \\

But let us see what kind of confusions may occur in such a process of setting up, and the
subsequent use of a mathematical model in physics. \\ \\

{\bf 2. Absolutising the Mathematical Model ...} \\

A rather trivial, even if not infrequent confusion is as follows. Once the mathematical model
is set up, it produces as consequences, and as is supposed to do of course, a number of
mathematical statements. And then, when interpreting them, it is often that one forgets the
essential difference between mathematics and physics. Namely, the validity of such
mathematical consequences is {\it not} absolute, but only conditional. And naturally, it is
conditional upon the extent to which the considerations of physics that led to the setting up
of the given mathematical model were, first, correct as such, and second, were correctly
expressed in mathematical terms. \\

Such a rather trivial confusion is illustrated in quantum mechanics, according to the
Copenhagen interpretation of its first von Neumann mathematical model, for instance, [vN].
Indeed, the celebrated paradox of Schroedinger's cat leads not a few to think that the state
of the cat is certainly undefined and somewhere vaguely between being alive and dead, right
until one opens the respective box and observes that state. This, of course, is quite the same
as thinking that, say, the Moon need not necessarily be there in its place in the sky, as long
as one is not looking at it, or the fall of a tree in a forest need not inevitably produce a
noise, unless someone who happens to be nearby would in fact hear it ... \\

Needless to say, interpretations of consequences of a mathematical model of physics are
welcome, being after all among the very reasons such mathematical models were set up in the
first place. And such interpretation should of course have a physical nature, since it is
physics itself which, before all other disciplines, is supposed to be the beneficiary of such
mathematical models. \\
What is inappropriate, however, is to claim an overriding validity for such physical
interpretations as coming from the fact that they are based on the given mathematical model,
and then do so forgetting that, in the first place, and as mentioned, the physical relevance
of that mathematical model is {\it not} supposed to be absolute. \\
Furthermore, a mathematical model is by {\it no} means supposed to deliver more than it was
included in its original setup. And then, the only new physical interpretations justified by
such a mathematical model are those which correspond to physical considerations that were {\it
implicitly}, rather than explicitly nevertheless included in it, or are merely the logical,
even if so far not yet known, consequences of physical considerations already included in the
mathematical model. \\

In the case of Schroedinger's cat, for instance, confusion arises, among others, from the fact
that the wave function $\psi$ of the respective system including the cat is a mathematical
entity in the respective von Neumann model, and as such it is subjected to the mathematical
rules of superposition, rules valid in absolutely any linear mathematical model, and not only
in that used by von Neumann in quantum  mechanics. Added to that comes the standard Born
interpretation which is but a bridge between mathematical probability and physics. \\

{\bf 3. Unwarranted Implications ...} \\

More subtle forms of misuse of mathematical models in physics may also occur. Let us give here
an example which applies both to special and general relativity. \\
As it happens, and is not often noted, light, which is well known to play a fundamental role
in both these theories, does {\it not} in any way whatsoever appear in their respective
mathematical models through any of its various physical properties, say, as a wave or as a
flux of photons, but only as a way of signalling between different points in space. And in
this regard, one only assumes the constancy in the velocity of that signalling, as well as
the fact that light is supposed to go, whatever that may mean, along geodesics, which in
special relativity are of course straight lines. \\
Indeed, even in general relativity, such a phenomenon as the bending of the trajectory of
light due to the gravitational effect of mass is {\it not} in any way a consequence of one or
another specific physical property of light, and as such, of its possible interaction with
mass, but solely of the curvature mass creates in space, curvature which leads to geodesics
that are no longer straight lines. \\
Consequently, any conclusion or interpretation based on the mathematical models of these two
theories and aimed at further elucidating the physical nature of light would be highly
questionable, and in fact, incorrect, since in the first place, nothing of the physical nature
of light, except as the mentioned way of signaling between different points in space, was
included in those models. \\
Of course, there is a clear general awareness that in the mathematical model of special
relativity there is nothing at all included about gravitation. Consequently, physicists do not
try to draw physical conclusions or make physical interpretations on gravitation, and do so
solely based on the mathematical model of special relativity. \\

On the other hand, in view of the mentioned fact that light is only included in these two
theories as a mere way to signal between different points in space, may allow a further
refinement of these theories, the moment one may try to incorporate into them {\it other}
important {\it physical} properties or aspects of light. \\ \\

{\bf 4. Two Frequent Forms of Misuse ...} \\

A third, and most frequent source of confusion, however, arises in the insufficiently proper
use of mathematical models of physics, and which happens as follows. There is a given
mathematical model of a certain theory of physics, and then, in the everyday use of this
mathematical model there is an ever ongoing forgetting of the essential difference between
such a mathematical model, and on the other hand, the theory of physics which it is supposed
to describe. \\

Amusingly however, this is not a simple passive forgetting, but on the contrary, a most
enthusiastically active one ! \\
And it manifests itself by ever more "mixing up" the given mathematics with an ongoing torrent
of "physical intuitions" ... \\

In this regard, one can distinguish between two frequent instances of such ongoing "mixing up"
in the inappropriate use of mathematical models in physics. \\ \\

{\bf 4.1. Improper "Tinkering" ...} \\

A first instance is when such an ever ongoing "mixing up" leads to no more than an ongoing
"tinkering" where one is no longer able to be clear enough where the mathematical precision,
so essential in physics, ends, and where one or another, and by its own very nature,
significantly less precise "physical intuition" is in fact replacing it ... \\
In this way, one is simply defeating the very reason mathematical modeling was introduced in
physics in the first place. \\
All that such "tinkering" may recall are those youngsters, passionate in their curiosity
about the inner workings of, say, computers, and who cannot stop "tinkering" with their own
computer, even if more often than not, the consequence is that such a computer becomes of not
much effective use ... \\

Needless to say, "physical intuitions" are absolutely essential in physics, just like all
other specialized intuitions are in the pursuit of their respective fields of science. \\
What is, however, simply defeating the whole purpose of mathematical modelling is to {\it
fail} to realize that, given a mathematical model, the appropriate way to use it, when wanting
to take into consideration "physical intuitions" that had not been included in it from the
very beginning, is to build from the scratch a whole new and refined mathematical model which,
this time, does include those "physical intuitions". \\
And in this regard, a very good example is that of the transition from special to general
relativity. Indeed, after setting up special relativity, Einstein never tried simply to flood
it with torrents of "physical intuition" regarding gravitation, and based on such intuitions,
to tinker with that theory until it may eventually be able to deal with gravitation itself.
Instead, he simply tried to start a rather radically new theory, and as is well known, went
through several variants of it, until in late 1915, he reached what he considered to be a good
formulation of what was to be called general relativity. \\ \\

{\bf 4.2. Unnecessary "Tinkering" ...} \\

The Bell inequalities give a clear, even if hardly known, example where in their mathematical
deduction {\it no} "physical intuition" of any quantum nature is in fact needed, yet the flood
of papers with such intuitions seems to be never ending, thus merely perpetuating the
respective lack of awareness and confusion ... \\

It is estimated, [A, p. 2], that by 1978, there were no less than over one million articles
and books on the celebrated EPR paper. \\
It is therefore not so surprising that a similar flood has emerged related to the Bell
inequalities, with one of the latest contributions being the paper [M]. \\

Most of such papers are, needless to say, written by physicists. Consequently, they tend to
exhibit, and be based upon a large variety of "physical intuitions". \\
Such a situation is not at all strange. After all, as known so well in other human endeavours,
be they philosophy, metaphysics, theology, art, or for that matter, economics and politics,
it is rather the rule than the exception that any given more important issue does inevitably
end up with a considerable number of views, interpretations and comments. \\

Furthermore, the EPR paper happened to highlight such deeply seated foundational issues in
quantum mechanics which, ever since its publication in 1935, could not be settled. \\
The Bell inequalities, on the other hand, even if having foundational implications of no
lesser importance, have the extraordinary feature to be at present easily testable by
effective physical experiments. And such tests, conducted in the early 1980s, showed clearly
their violation in quantum mechanical contexts. Therefore, unlike with the issues raised in
the EPR paper, the Bell inequalities tend to provoke discussions focused more on their quantum
mechanical connection and relevance, than on the status as a whole of quantum mechanics, or of
its various interpretations, among them, the Copenhagen one. \\

And yet, the flood of studies and comments on the Bell inequalities keeps going on ... \\
And so far, it only seems to increase a general confusion and misunderstanding, and do so more
than contributing to clarity. Indeed, as shown back in 1989 by I Pitowsky, [P], the Bell
inequalities happen to belong to a more general family of inequalities in {\it elementary
probability theory}, presented as long ago as in chapter 21 of the celebrated 1854 book of
George Boole, "The Laws of Thought",[B, B1]. Therefore, they cannot have much to do with
quanta. \\

What seems to be in this case the main underlying dynamics is therefore rather different from
that in the case of the EPR paper. \\

And as shown in [P], [R], this dynamics comes from the fact that in such studies and comments
{\it unnecessary} and simply {\it superfluous} "physical intuition" is being injected. \\
Indeed, as mentioned, the Bell inequalities are related to the family of inequalities in
elementary probability theory presented, as long ago as during Victorian times, namely, in
chapter 21 of the celebrated 1854 book of George Boole, "The Laws of Thought", [B], [B1]. \\
Consequently, they cannot possibly have much to do with quantum theory, hidden variables,
local hidden variables, and so on ... \\

As a morale, one should recall the important methodological advise of good old lawyers that,
when dealing with an issue properly, two sharply conflicting avenues should be pursued at the
same time : to associate with that issue everything which is relevant, and also, to
disassociate from that issue all that is not relevant. \\

As it happens, however, with "physical intuition", it appears that in its practice it is not
so easy to follow that methodological advise ... \\

And the flood of papers on the Bell inequalities is as clear an example of such a failure, as
any, even if it may take some time until it would be recognized as such ... \\ \\

{\bf 5. What Mathematical Models Could Do for Physics ... } \\

There is still another aspect of the less than proper use of mathematics as a whole within
physics. Namely, we refer here not so much to the setting up of one or another specific
mathematical model for a certain theory of physics, but rather to what mathematics may
inspire and support in the process of creative imagination of physicists, which is so
essential for their "physical intuitions". \\
However, from the start, one has to note that here one may face a situation which may indeed
be far more difficult to deal with, since it would require no less than a truly up to date
awareness on the part of physicists about the newest possibilities offered by state of the
art mathematics. And as mentioned, here we are not talking only about the mathematical
modelling  physicists become involved in, but about the very {\it freedom of imagination} of
their "physical intuitions", imagination which such mathematics may suggest, inspire, allow,
offer and also support. \\

Here we mention two such instances, and regrettably they still seem much beyond the pale for
most physicists, even if they happen to involve some modern mathematics which for longer has
no more been state of the art, but rather well known and established, thus could have been
used in physics for quite some time by now. \\

One of them, already with some awareness in certain circles of physicists, is about what {\it
scalars} could and should be used in physics, see [R3] and the literature cited there. \\
In this regard, one can recall that it was in the study of electricity in the second half of
the 1800s, when the use of {\it complex scalars} got firmly established in the mathematical
modelling of physics. \\
Fortunately therefore, there was no resistance to be encountered among physicists by the time
of Schroedinger's use in the mid 1920s of complex scalars in his celebrated quantum wave
equation, and thus of von Neumann's subsequent use of such scalars in his first mathematical
model of quantum mechanics, [vN]. \\

The second example regards the possible use of {\it non-Archimedean} structures which, among
others, may be appropriate, for instance, for the mathematical modelling of space-time,
[R1,R2]. And here one should note that, no matter how successful such structures have already
proved themselves in a number of important branches of mathematics, among them, the solution
of very large classes of nonlinear partial differential equations, the reluctance to consider
them at all has so far marked the physics community. \\
And amusingly, such a reluctance amounts to nothing else by a {\it self-censorship}, imposed
by that community upon itself, even if imposed by default, as can be seen already from the
rather particular example considered in [R1], and which can hint to the extraordinary realms
of freedom for the creativity of imagination of "physical intuition". \\


\begin{thebibliography} {99}
\bibitem{} [M] Matzkin A : Is Bell's theorem relevant to quantum mechanics?
arxiv:0802.0613v2
\bibitem{} [R3] Rosinger E E : How far should the Principle of \\ Relativity go ?
arxiv:0710.0226v3
\end{thebibliography}
\end{document}